\documentclass[12pt]{iopart}

\usepackage{iopams,graphicx}
\begin{document}
\title[Transport of the repulsive Bose-Einstein condensate ...]
{Transport of the repulsive Bose-Einstein condensate in a
double-well trap: interaction impact and relation to  Josephson
effect}
\author{V O Nesterenko$^1$, A N  Novikov$^1$ and E Suraud$^2$}
\address{$^1$Bogoliubov Laboratory of Theoretical Physics, Joint
Institute for Nuclear Research, Dubna, Moscow region, 141980,
Russia} \ead{nester@theor.jinr.ru}
\address{$^2$Laboratoire de Physique Quantique, Universit\'e Paul
Sabatier, 118 Route de Narbonne, 31062 cedex, Toulouse, France}

\begin{abstract}
Two aspects of the transport of the repulsive Bose-Einstein
condensate (BEC) in a double-well trap are inspected: impact of
the interatomic interaction and analogy to the Josephson effect.
The analysis employs a numerical solution of 3D time-dependent
Gross-Pitaevskii equation for a total order parameter covering all
the trap. The population transfer is driven by a time-dependent
shift of a barrier separating the left and right wells. Sharp and
soft profiles of the barrier velocity are tested. Evolution of the
relevant characteristics, involving phase differences and
currents, is inspected. It is shown that the repulsive interaction
substantially supports the transfer making it possible i) in a
wide velocity interval and ii) three orders of magnitude faster
than in the ideal BEC. The transport can be approximately treated
as the d.c. Josephson effect. A dual origin of the critical
barrier velocity (break of adiabatic following and d.c.-a.c.
transition) is discussed. Following the calculations, robustness
of the transport (d.c.) crucially depends on the interaction and
barrier velocity profile. Only soft profiles which minimize
undesirable dipole oscillations are acceptable.
\end{abstract}

\maketitle

\section{Introduction}

The population transfer is a typical problem met in various
branches of physics (ultracold gases and condensates
\cite{Petrick_Smith_02,Pit_Str_03,Dalfovo_99,Leggett_01,Yuk_rew_01,Gati_07,Torr_12_rev,Dal_11,
Yuk_13}, atomic and molecular physics \cite{Kral}, etc.). The
problem is easily solvable, if it is linear and accepts an
adiabatic evolution, see e.g. the Landau-Zener scenario
\cite{Lan,Zen}. However, if there are significant nonlinear
effects or/and we need a rapid but robust transfer, the problem
becomes nontrivial, like e.g. in the irreversible nonlinear
transport (NLT) of Bose-Einstein condensate (BEC) in multi-well
traps \cite{Nest_LP_09,Nest_JPB_09}. The trapped BEC is especially
suited for investigation of nonlinear transport  because BECs
features, including the interaction-induced nonlinearity, can be
precisely controlled and manipulated. Besides,  by driving the
trap parameters one can  simulate various transport protocols.

Despite numerous experimental and theoretical studies (see early
\cite{Petrick_Smith_02,Pit_Str_03,Dalfovo_99,Leggett_01,Yuk_rew_01,
Gati_07} and recent \cite{Torr_12_rev,Ruschhaupt_NJP_12} reviews),
some important NLT features yet poorly understood. In particular,
it is not well established in which cases the nonlinearity favors
the  transport and how essential is the effect.

In the present study, we address these general questions for a
typical NLT scheme: an external Bose Josephson junction (EBJJ)
produced in a double-well trap. Here the left and right BEC
fractions are coupled through the barrier separating the tap. The
nonlinear effects are caused by interaction between BEC atoms. The
NLT is a population inversion driven by converting the trap from
initial to final (opposite) asymmetric configurations. Nowadays
such INTL is a routine experimental operation which can be
produced by various methods: from familiar Rabi oscillations
($\pi$ pulses) \cite{Allen_book_87} and (quasi)adiabatic
population transfer \cite{Lan,Zen,Nest_JPB_09} to modern
shortcut-to-adiabaticity methods (see review \cite{Torr_12_rev}
and particular relevant options \cite{Berry_09,JD10,JD13})
promising a fast and robust population inversion. The goal of the
present study is to use this simple operation for exploration of:
i) strong nonlinear effects predicted for this configuration
within a simple two-mode model \cite{Nest_JPB_09}, ii) analogy
between NLT and d.c. (direct current) Josephson effect in
superconductors \cite{Jo62}, predicted
\cite{Sme97,Za98,Rag99,Gi00,Meier01,Sak02} and observed
\cite{Le07} in EBJJ.

For this purpose,  the three-dimensional (3D) time-dependent
Gross-Pitaevskii equation (GPE) \cite{GPE} for the total order
parameter covering both left and right parts of the condensate in
a double-well trap is numerically solved. The calculations are
free from the two-mode approximation (TMA) \cite{two-mode} and
other simplifications used in our previous estimations
\cite{Nest_JPB_09}. Furthermore, our study  closely follows
conditions and parameters of Heidelberg's experiments
\cite{Albiez_exp_PRL_05,Gati_APB_06}, thus providing atypical but
realistic picture. The population transfer is determined by a
time-dependent barrier shift driving the system between initial
and final asymmetric configurations. This technique allows to
reach simultaneously two aims: i) exercise a generalized
Landau-Zener/Rosen-Zener transport protocol implemented in our
previous study \cite{Nest_JPB_09} and ii) simulate an external
current required for generation of Josephson d.c. in EBJJ
\cite{Gi00}. To highlight nonlinear effects, the dynamics of ideal
and repulsive BEC is compared.

In our previous TMA study, a strong support of the transport by
the repulsive interaction was found \cite{Nest_JPB_09}. It was
shown that the interaction leads to a wide range (plateau) of the
process rates, where a complete (quasi)adiabatic transport is
realized. In the present study, we test these results within a
more realistic model beyond the TMA. The scale of the nonlinear
effects is estimated for the particular Heidelberg setup
\cite{Albiez_exp_PRL_05,Gati_APB_06}. It is shown that the
repulsive BEC can be transferred by 3 orders of magnitude faster
than the ideal condensate. A pollution of NLT by dipole
oscillations is estimated and a smooth velocity regime moderating
this problem  is proposed.

In the second part of our exploration, the NLT
is compared with the Josephson d.c. and a.c. effects \cite{Jo62}
represented for BEC by equations \cite{Gi00}
\begin{equation}\label{Jo}
  I=I_0\sin(\theta), \quad \dot{\theta}=\frac{\Delta\mu}{h} ,
\end{equation}
where $I$ is the supercurrent, $I_0$ is its critical value and
$\Delta\mu$ is the difference between chemical potentials of the
wells. As predicted \cite{Gi00} and then experimentally observed
\cite{Le07}, the  d.c. can be generated in EBJJ by an adiabatic
movement of the barrier across the trap with a constant velocity,
thus simulating the driving current. The shift can drive the trap
from asymmetric to symmetric configuration \cite{Le07} or vice
versa \cite{Jo62}. The adiabatic evolution assumes that the system
change is so slow that tunneling of atoms between the wells is
sufficient to lock $\Delta\mu$ to zero. When the shift is over, we
get the Josephson d.c. $I$ driven by the phase difference
$\theta$. The critical current $I_0$ should be proportional to the
critical velocity $v^{\rm{crit}}$ of the barrier shift. Above this
velocity, the adiabatic flow breaks down, the nonzero $\Delta\mu$
develops, and the process becomes of a.c. character with $I=I_0
\sin (\Delta\mu t/\hbar)$ \cite{Gi00,Meier01,Sak02}.

It is easy to see that this scenario corresponds to an adiabatic
NLT described within the TMA in our previous study
\cite{Nest_JPB_09}. The plateau in the transport rates
\cite{Nest_JPB_09} is just the region $I < I_0$ where the
adiabatic evolution takes place. The critical rate
\cite{Nest_JPB_09} marking the break of the adiabatic transport
seems to correspond to $v^{\rm{crit}}$ and $I_0$ in
\cite{Gi00,Le07}. The analogy should take place despite the
population transfer in \cite{Nest_JPB_09} is driven not by the
barrier shift but by another technique generalizing Landau-Zener
and Rosen-Zener schemes. Both scenarios have to be physically
similar since they satisfy the principle requirements: weak
coupling, inherent phase difference, and adiabatic evolution.

In the present study, we continue analysis of d.c. and a.c. in
EBJJ but now with the accent to nonlinear effects. As compared to
the previous studies \cite{Za98,Rag99,Gi00,Meier01,Sak02} which
were limited to inspection of the population imbalance $z$  and
chemical potential difference $\Delta\mu$, we also scrutinize the
evolution of the phase difference $\theta$, a principle factor of
the Josephson dynamics. In particular, we provide a detailed
analysis of $\theta$ near $v^{\rm{crit}}$. Also, a pollution
effect of the dipole oscillations is estimated. It is shown that
the constant barrier velocity \cite{Gi00} results in strong
oscillations which greatly smear the process and complicate the
analysis. Thus, a soft velocity profile is proposed to circumvent
this trouble. It is shown that the repulsive interaction and soft
velocity profile make the NLT (and d.c./a.c.) much more suitable
for the analysis and experimental observation.

Note that last years EBJJ is widely used in diverse actual areas
(shortcuts to adiabaticity and optimal control
\cite{Torr_12_rev,Ruschhaupt_NJP_12}, spin squeezing, entanglement
and quantum metrology \cite{Gross_JPB_12,Wer_JPB_07}, Josephson
dynamics in spin-orbit BEC \cite{Gar14}, etc). At the same time,
investigations of  d.c./a.c. regimes in EBJJ are yet sparse
\cite{Rad10}, despite interesting flaring similarity of d.c. with
adiabatic population transfer scenarios. The present detailed
study of a.c./d.c. in a double-well trap aims to  supply partly
this gap.

 The paper is organized as follows.  The theory and calculation framework
are outlined in Sec. \ref{sec:calc_scheme}. The results are
discussed in Sec. \ref{sec:results}. The summary is given in Sec.
\ref{sec:summary}.

\section{Calculation scheme}
\label{sec:calc_scheme}

\subsection{Trap setup and well populations}

The calculations are performed  within the 3D time-dependent
Gross-Pitaevskii equation (GPE) \cite{GPE}
\begin{equation}\label{GPE}
 i\hbar\frac{\partial\Psi}{\partial t}({\bf r},t) =
[-\frac{\hbar^2}{2m}\nabla^2 + V({\bf{r}},t) + g_0|\Psi({\bf
r},t)|^2]\Psi({\bf r},t)
\end{equation}
for the total order parameter $\Psi({\bf r},t)$ describing BEC in
both left and right wells of the trap. Here $g_0=4\pi\hbar^2 a_s
/m$ is the interaction parameter, $a_s$ is the scattering length,
and $m$ is the atomic mass. The trap potential
\begin{eqnarray} \label{trap_pot}
V({\bf{r}},t)&=& V_{\rm{con}}({\bf{r}})+V_{\rm{bar}}(x,t)
 \\
 &=&\frac{m}{2}(\omega^2_x x^2+\omega^2_y
y^2+\omega^2_z z^2) \nonumber
\\&+& V_0 \cos^2(\pi (x-x_0(t))/q_0) \nonumber
\end{eqnarray}
includes the anisotropic harmonic confinement and the barrier in
$x$-direction, whose position is driven by the control function
$x_0(t)$ \cite{Gi00,Sak02}; $V_0$ is the barrier height and $q_0$
determines the barrier width.

Following conditions of the Heidelberg experiment
\cite{Albiez_exp_PRL_05,Gati_APB_06} (where Josephson oscillations
(JO) and macroscopic quantum self-trapping (MQST) have been
observed), we consider BEC of N=1000 $^{87}$Rb atoms with
$a_s=5.75$ nm. The trap frequencies are $\omega_x=2\pi\times 78$
Hz, $\omega_y=2\pi\times 66$ Hz, $\omega_z=2\pi\times 90$ Hz, i.e.
$\omega_y+\omega_z=2\omega_x$. The barrier parameters are
$V_0=420\times h$ Hz and  $q_0=5.2 \; \mu$m. For the symmetric
trap ($x_0$(t)=0), the distance between the centers of the left
and right wells is $d=$4.4 \;$\mu$m. This setup has been earlier
used in our exploration of JO/MQST in a weak and strong coupling
\cite{Nest_JPB_12}. It corresponds to so called Josephson
(classical) regime when quantum fluctuations of both population
imbalance and phase difference are not essential.

The static solutions of GPE are found within the damped gradient
method \cite{DGM} while the time evolution is computed within the
time-splitting technique \cite{TSM}. The total order parameter
$\Psi({\bf r},t)$ is determined in a 3D cartesian grid. The
conservation of the number of atoms, $\int^{-\infty}_{+\infty}dr^3
|\Psi({\bf r},t)|^2 =N$, is directly fulfilled by using an
explicit unitary propagator. No  time-space factorization of the
order parameter is implemented. The conservation of the total
energy $E$ is controlled.

The populations of the left (L) and right (R) wells are computed
as
\begin{equation}\label{N_LR}
N_{j}(t)=\int^{+\infty}_{-\infty}dr^3 |\Psi_{j}({\bf r},t)|^2,
\end{equation}
with $j = L, R$, $\Psi_{L}({\bf r},t)=\Psi(x\le x_0(t),y,z,t)$,
$\Psi_{R}({\bf r},t)=\Psi(x\ge x_0(t),y,z,t)$ and
$N_L(t)+N_R(t)=N$. The normalized population imbalance is
\begin{equation}
z(t) = (N_L(t)-N_R(t))/N .
\end{equation}

The NLT to be considered means that initial (t=0) BEC populations
 $N_L(0) > N_R(0)$ are inverted during the time T to the final
populations $N_L(T) < N_R(T)$ where $N_L(T)=N_R(0)$ and
$N_R(T)=N_L(0)$. The initial stationary {\it asymmetric} BEC state
is produced by adjusting the barrier right-shift ($x(0)> 0$) so as
to provide the required initial populations $N_L(0)$ and $N_R(0)$.
The NLT is achieved by a barrier left shift from $x(0)$ to
$x(T)=-x(0)$ with the shift velocity $v(t)$. Thus the trap
asymmetry is changed to the opposite one.

Two velocity time profiles are used: i) the sharp rectangular one
with the constant $v_c(t)=v^c_0$ at $0 < t < T$ and $v_c(t) =$ 0
beyond the transfer time, and ii) the soft one $v_s(t)=v^s_0
\cos^2(\frac{\pi}{2}+\frac{\pi t}{T})$ with $v_s(0)=v_s(T) \sim 0$
and  $v_s(T/2)=v^s_0$. For the total barrier shift $D=2 x(0)$ in
the inversion process of duration T, the velocity amplitudes are
$v^c_0=D/T$ and $v^s_0=2D/T$. The average velocities
$v_a=D/T$ are:
\begin{equation}
v^c_a=v^c_0, \qquad  v^s_a=v^s_0/2 .
\end{equation}
The constant profile $v_c(t)$ was used in
previous studies \cite{Gi00,Sak02}. It sharply changes from 0 to
$v^c_0$ at t=0 and back at t=T and, in this sense, is not
adiabatic. As shown below, the sharp changes cause undesirable
dipole oscillations which can significantly pollute the population
transfer. The second profile $v_s(t)$ is softer and thus closer to
the adiabatic evolution.

The NLT quality is characterized by its completeness
$P=-z(T)/z(0)$ (the ratio of the final and initial population
imbalance) and noise $n=A_d/N$ where $A_d$ is amplitude of dipole
oscillations in the final state, i.e.
$A_d=\rm{max}\{N_{L,R}\}-\rm{min}\{N_{L,R}\}$ for $t>T$.

Note that previous studies used  3D \cite{Gi00} and 1D
\cite{Sak02} numerical time-dependent GPE simulations as well.

\subsection{Phases}

The phases $\phi_{j}(t)$ of the left and right BEC fractions are
defined as \cite{Nest_JPB_12}
\begin{equation}\label{phi_LR0}
\varphi_{j}(t)=\arctan \frac{\gamma_j(t)}{\zeta_j(t)}
\end{equation}
with the averages
\begin{eqnarray}\label{gamma}
\varsigma_j (t)=\frac{1}{N_j}\int^{+\infty}_{-\infty}dr^3
\mathrm{Im}(\Psi_j({\bf r},t))|\Psi_j({\bf r},t)|^2 \; ,
\\ \label{zeta}
\chi_j (t)=\frac{1}{N_j}\int^{+\infty}_{-\infty}dr^3
\mathrm{Re}(\Psi_j({\bf r},t))|\Psi_j({\bf r},t)|^2 .
\end{eqnarray}
Since computation of the phase time evolution through $\arctan$
may be cumbersome, we use (\ref{phi_LR0}) only for the static case
while the time evolution is calculated through the phase
increments $\varphi_{j}(t+\delta t)\approx \varphi_{j}(t)+\delta
\varphi_{j}(t)$ for a small time step $\delta t$.  Namely, we use
\begin{equation}\label{dphi}
\delta\varphi_{j}(t)=
\sqrt{\frac{[\delta\varsigma_j(t)]^2+[\delta\chi_j (t)]^2}
{\varsigma_j^2(t+\delta t)+\chi_j^2(t+\delta t)}}
\end{equation}
with $\delta\varsigma_j(t)=\varsigma_j(t+\delta
t)-\varsigma_j(t)$, $\delta\chi_j(t)=\chi_j(t+\delta
t)-\chi_j(t)$. The phase difference is
\begin{equation}\label{phi_LR0}
\theta(t)=\varphi_{R}(t)-\varphi_{L}(t) .
\end{equation}

\subsection{Energy estimations}

 To discriminate weak and strong couplings between BEC
fractions, it is instructive to compare the energy of the occupied
state
with the barrier height $V_0$. Since the barrier takes place in
x-direction, only the part of the ground state energy in the same
direction is relevant. In the linear case ($g_0$=0), the total
ground state energy reads as in anisotropic harmonic oscillator,
$\mu_0=\mu_{x0}+\mu_{yo}+\mu_{z0}$, and its relevant x-part is
\begin{equation}\label{mu_x0}
\mu_{x0} = \mu_0 -
\frac{\hbar}{2}(\omega_y+\omega_z)=\mu_0-\hbar\omega_x=\alpha
\mu_0
\end{equation}
where the relation $\omega_y+\omega_z=2\omega_x$
\cite{Albiez_exp_PRL_05,Gati_APB_06} is used. The numerical GPE
estimation gives $\alpha =\mu_{x0}/ \mu_0 \approx 3/4$
\cite{Nest_JPB_12}.

In the nonlinear case ($g_0 \ne$ 0), the estimation of $\mu_x$  is
straightforward for 1D system but demanding for 3D case considered
here. So we use the simple ansatz
\begin{equation}\label{mu_x}
\mu_x = \alpha \mu
\end{equation}
where $\mu$ is the total {\it nonlinear} ground state energy and
$\alpha \approx 3/4$ as in the linear case. This phenomenological
relation was shown to be accurate in investigation of the
evolution of JO/MQST dynamics under the transition from a weak to
a strong coupling \cite{Nest_JPB_12}. In this study, it is used
only for illustrative aims, namely for the comparison with  the
barrier height $V_0$ in Fig. 1.

The energies $\mu_0$ and $\mu$ can be treated as chemical
potentials in the Josephson setup
\cite{Petrick_Smith_02,Pit_Str_03,Dalfovo_99,Leggett_01}. In the
rapid evolution of the system, initiated by the barrier shift, the
difference between chemical potentials of the left and right
wells, $\Delta\mu=\mu_L -\mu_R$, can be created \cite{Gi00}. In
NLT, $\Delta\mu$ can be estimated through $\dot{\theta}$, see Eq.
(\ref{Jo}).

\begin{figure*}
\begin{center}
\includegraphics[width=11cm]{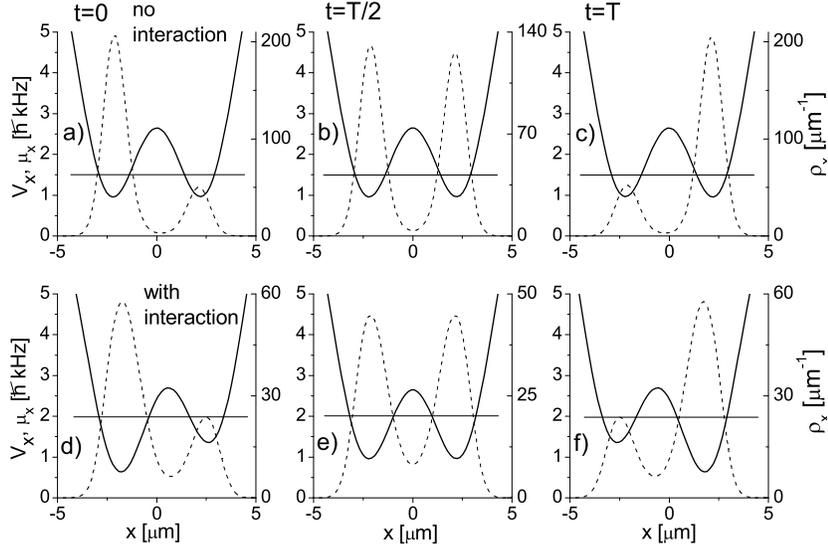}
\end{center}
\caption{The double-well trap potential $V_x(x)$ (bold curve), BEC
density $\rho_x(x)$ (dash curve), estimated ground state energy
$\mu_x$ (solid strait line) at the initial (t=0), intermediate
(t=T/2) and final inverse (t=T) states of the adiabatic inversion,
calculated without (upper plots) and with (bottom plots) the
repulsive interaction between BEC atoms. In both cases, the
initial populations of the left and right wells are $N_L(0)$=800
and $N_R(0)$=200.}
\end{figure*}

\subsection{Josephson current}

The Josephson current is defined as
\begin{equation}\label{I}
I(t)=-\frac{\dot{z}(t)}{2}=-\frac{\dot{N}_L(t)}{N}=\frac{\dot{N}_R(t)}{N}.
\end{equation}
This explicit current may be compared to an approximate one
\begin{equation}\label{I_approx}
\tilde{I}(t)=I_0
\sqrt{1-z(t)^2}\sin \theta(t)
\end{equation}
following from the first of the GPE-TMA equations
\cite{Nest_JPB_09,Sme97,Rag99,two-mode}:
\begin{eqnarray}\label{z_2m}
\dot{z}&=&-2K\sqrt{1-z^2}\sin \theta ,
\\
\label{theta_2m}
\dot{\theta}&=&\frac{\Delta\mu}{2}+K\frac{z}{\sqrt{1-z^2}} \cos
\theta +\frac{NU}{2}z .
\end{eqnarray}
Here  $I_0$ is the EBJJ critical current, $K$ is the coupling
between BEC fractions through the barrier, $U$ is the interaction
between BEC atoms inside the trap wells. In the TMA, we have
$I_0=2K$. Eqs. (\ref{z_2m})-(\ref{theta_2m}) are mathematically
similar to those for resonantly generated coherent modes
\cite{Yuk_04}. What is important for our aims, Eqs.
(\ref{z_2m})-(\ref{theta_2m}) remind the Josephson equations
(\ref{Jo}).

In our study, we get the population imbalance $z(t)$, phase
difference $\theta (t)$ and currents $I(t)$ and $\tilde{I}/I_0$
not from (\ref{z_2m})-(\ref{theta_2m}) but from a direct solution
of the GPE (\ref{GPE}). Then the comparison of the explicit
(\ref{I}) and approximate (\ref{I_approx}) currents at the
reasonable point, say at $t=T/2$, allows to estimate the critical
current $I_0$.

\section{Results and discussion}
\label{sec:results}

\subsection{Confinement, density and chemical potential}

Figure 1 exhibits  the trap potential in x-direction,
\begin{equation}
V_x(x,t)=\frac{m}{2}\omega^2_x x^2 + V_0 \cos^2(\pi
(x-x_0(t))/q_0) ,
\end{equation}
 calculated for the initial t=0, intermediate t=T/2 and final t=T
times of the inversion process.
For the same times, the BEC density profile in
x-direction,
\begin{equation}
\rho (x,t)=\int^{+\infty}_{-\infty} dy dz |\Psi (x,y,z,t)|^2 ,
\end{equation}
obtained for an adiabatic inversion of a long duration T is shown.
The ideal and repulsive BECs with N=1000 atoms are considered.
Following the plots a) and d), the initial populations of the left
and right wells are $N_L(0)$=800 and $N_R(0)$=200 with the
population imbalance  $z(0)$=0.6.    An adiabatic evolution
provides a robust population inversion to final state with
$N_L(T)$=200, $N_R(T)$=800 and  $z(T)$=-0.6. At the intermediate
time t=T/2, the trap and populations are symmetric. The initial
state is stationary by construction. The intermediate and final
states, being obtained adiabatically, can be also treated as
stationary.
\begin{figure}[h]
\begin{center}
\includegraphics[width=7cm]{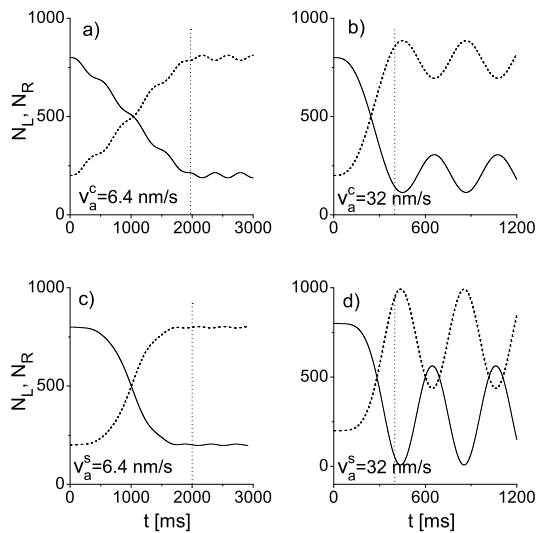}
\end{center}
\caption{Evolution of populations $N_L(t)$ (solid curve) and
$N_R(t)$ (dash curve) in the {\it ideal} (no interaction) BEC,
calculated at the initial $N_L$(0)=800, $N_R$(0)=200 and
$z$(0)=0.6. Durations of the barrier shifts $T$=2 s (a,c) and 0.40
s (b,d) are marked by vertical dotted lines. The transfers with
the constant  $v^c(t)$ (a,b) and soft $v_s(t)$
(c,d) velocity profiles are considered (with the same average
velocities as indicated).}
\end{figure}

Upper plots of Fig. 1 show that for getting  the initial
$z(0)$=0.6 in the ideal BEC, a small trap asymmetry with
$x_0(0)$=0.0064 $\mu$m is sufficient. The overlap of the left and
right parts of the condensate at the center of the trap is very
small. The chemical potential $\mu_{x}$ from (\ref{mu_x0}) lies
much below the barrier top. The energy difference between the
ground and first excited states at the mid of the transfer (plot
b)) is $\Delta \mu(T/2)/h$ = 5 Hz, i.e. much smaller than the well
depths and trap frequencies. Altogether all these factors indicate
a weak coupling case.

For the repulsive BEC (bottom plots), the initial  $N_L(0)$=800
and $N_R(0)$=200 are obtained at a much larger asymmetry with
$x_0(0)$=0.5 $\mu$m. The energy splitting $\Delta \mu(T/2)/h$
reaches 36 Hz. The repulsive interaction significantly increases
the chemical potential $\mu_x$ (\ref{mu_x}) and thus the coupling
between the left and right BEC fractions. In this case, to get the
initial {\it stationary} population imbalance $z(0)$=0.6, one
should appreciably  weaken the coupling by the corresponding
increasing the asymmetry. As compared to the ideal BEC, the
repulsive condensate has wider density bumps which significantly
overlap at the center of the trap. The coupling between the left
and right BEC fractions is not weak anymore, though the NLT
considered below is yet realized through tunneling.

\subsection{Linear and nonlinear dynamics}

Some examples of the time evolution of the populations
$N_{L,R}(t)$ in the ideal BEC are given in Fig. 2. The evolution
is driven by the barrier shift with the rectangular $v_c(t)$
(upper plot) and soft  $v_s(t)$ (bottom plots) velocity profiles.
In both cases, the same average velocities are used. The total
barrier shift is D=12.8 nm. It is seen that, at low (adiabatic)
velocities corresponding to a long time T=2 s (plots a),c)), we
get a robust population inversion. The final state is about
stationary for $v_s(t)$ and somewhat spoiled by dipole
oscillations for $v_c(t)$. The latter is caused by the sharp
change of $v_c(t)$  at the beginning and end of the process. In
this sense, the $v_s(t)$-transfer  is softer and more adiabatic.
Following plots b),d), the inversion becomes worse or even breaks
down at high velocities.

\begin{figure}[h]
\begin{center}
\includegraphics[width=7cm]{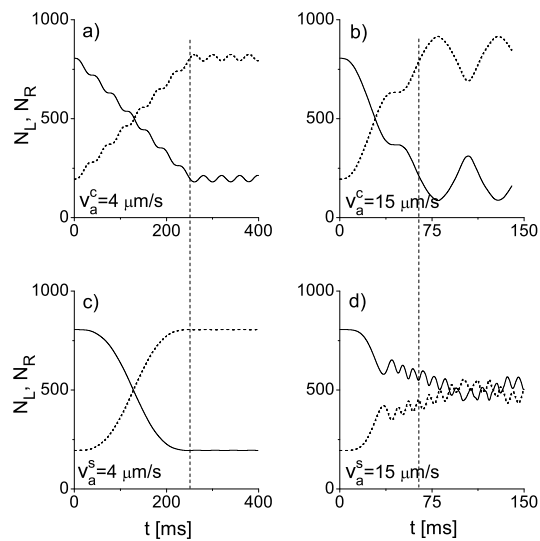}
\end{center}
\caption{The same as in Fig. 2 but for the repulsive BEC. The
barrier shift durations  are $T$=0.25 s (a,c) and 0.067 s (b,d).
}
\end{figure}
In Fig. 3,  similar examples are given for the repulsive BEC. At
first glance, the non-linear evolution resembles the linear one in
Fig. 2. Like in the linear case,  a slow transfer (plots a,c)
results in a robust NLT while a faster process (plots b,d) spoils
the final state by dipole oscillations (b) or even breaks the
inversion at all (d). However, the nonlinearity essentially
changes rates of the process. The robust NLT are produced for
larger barrier shifts (1 $\mu$m instead of 0.013 $\mu$m), for much
shorter times (T=250 ms instead of T=1800 ms for ideal BEC), and
with much faster velocities ($\mu$m/s instead of nm/s). The
velocities are three order of magnitude higher (!) than in the
linear case. The repulsive interaction greatly favors the
population inversion (the transfer parameters become more
comfortable for the experiment) and  the effect is indeed huge.
The  reason is in the growth of the chemical potential $\mu$,
caused by the repulsive interaction. This leads to a dramatic
increase of the barrier penetrability. The coupling between BEC
fractions becomes strong and the inversion is realized much
faster.

A more general information on NLT and is presented in Figs. 4 and
5 where the completeness $P$ and noise $n$ of the inversion are
given for a wide range of velocity amplitudes. In Fig. 4, the
sharp velocity profile $v_c(t)$ is used. Following the plots a,c)
for the ideal BEC, a complete inversion ($P$=1) takes place only
at a small velocity $v^c_0 <$ 0.04 $\mu$m/s. The inversion is
somewhat spoiled by a noise $n=$ 0.02 - 0.04 which weakens with
decreasing the velocity. For $v^c_0 >$ 0.04 $\mu$m/s, we see a
gradual destruction of the inversion, accompanied by an enhanced
noise. For even larger velocities, the inversion breaks down ($P
\to $ 0) and the final state is characterized by strong Rabi
oscillations ($n \to$ 0.4). The oscillations are caused by the
instant change of the velocity from zero to $v^c_0$ at t=0 and
back at t=T.
\begin{figure}[h]
\begin{center}
\includegraphics[width=8cm]{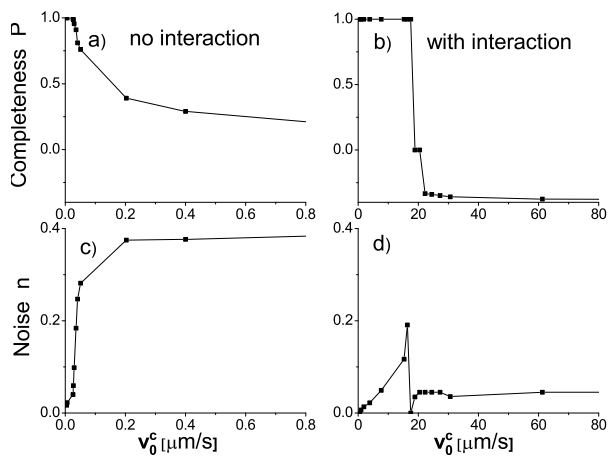}
\end{center}
 \caption{Completeness a)-b) and noise c)-d) of the
population inversion for BEC without (left plots) and with (right
plots) repulsive interaction versus the constant velocity
$\bar{v}^c_a=v^c_0$.}
\end{figure}
\begin{figure}[h]
\begin{center}
\includegraphics[width=8cm]{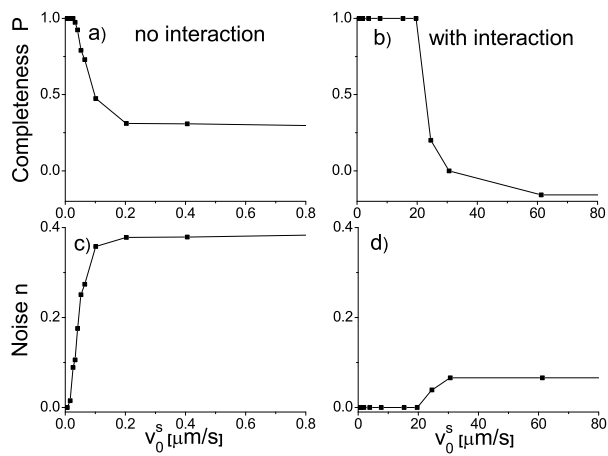}
\end{center}
\caption{The same as in Fig. 4 but as a function of the {\it maximal}
velocity $v^s_0$ (profile $v_s(t)$).}
\end{figure}

Following Fig. 4 b,d), inclusion of the repulsive interaction
dramatically changes the results. There appears a wide plateau, $0
< v^c_0 \le$ 19 $\mu$m/s (with the critical velocity $v^{\rm crit}
\approx$ 19  $\mu$m/s), where the inversion  is about complete ($P
\approx $ 1). As mentioned above, the repulsive interaction allows
to get the inversion three orders of magnitude faster than for the
ideal BEC. These findings are in accordance with  our previous
results for NLT, obtained within the simplified TMA model
\cite{Nest_JPB_09}.

Note that in the ideal and repulsive BEC the inversion breaks down
by different ways. While in the linear case the transfer
completeness $P$ tends not to zero, in the repulsive BEC it
becomes negative, $P \approx $-0.7. The later means that $z(0)$
and $z(T)$ have the same sign, i.e. the process results only in a
partial population transfer, keeping the initial inequality $N_L >
N_R$  at t=T.

In Figure 5, the similar analysis is done for the softer velocity
profile $v_s(t)$. Note that, as compared to Figs. 2 and 3, here we
use not the average $v^s_a $ but maximal velocity $v^s_0=2v^s_a$.
The results are generally similar to those in Fig. 4. However, in
the repulsive BEC (Fig. 5d), the process below the critical
velocity is much less noised than in the previous $v_c(t)$ case.
So, as might be expected, the softer (more adiabatic) profile
$v_s(t)$ leads to a more robust inversion than the sharp profile
$v_c(t)$.

In the repulsive BEC, the critical velocities for both profiles,
$v_c^{\rm crit}\approx 19 \mu$m/s and $v_s^{\rm crit}\approx 22
\mu$m/s, are rather similar. Note that these upper limits concern
{\it maximal} (not average for $v_s(t)$) velocities. The physical
sense of the critical velocity  is simple: destruction of the
adiabatic following \cite{Nest_JPB_09}. Namely, if the
 system is transformed slowly, then the tunneling suffices
to arrange BEC distribution in accordance to the transformation.
Thus we gain the adiabatic NLT. However, at a critical velocity,
the transformation becomes too fast and the efficient adiabatic
transfer (transport) breaks down. This argument is partly
confirmed by the fact that $v^c_{\rm crit}< v^s_{\rm crit}$, i.e.
the softer velocity profile leads to a bigger critical velocity.
More insight into the nature of $v^{\rm crit}$ can be reached by
treating NLT in terms of Josephson direct and alternating currents
\cite{Gi00}, see the next subsection. Then $v^{\rm crit}$ is
associated to the critical current manifesting the d.c. $\to$ a.c.
transition. However,  d.c. also assumes an adiabatic following and
so does not contradict the adiabatic arguments of Ref.
\cite{Nest_JPB_09}.
\begin{figure}[h]
\begin{center}
\includegraphics[width=7cm]{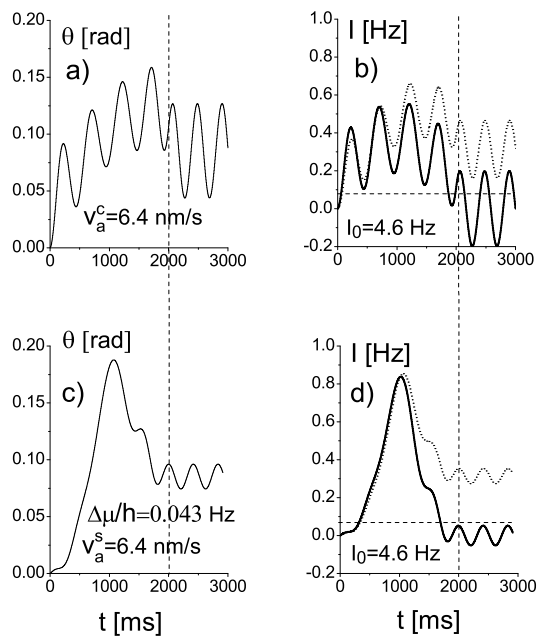}
\end{center}
\caption{Evolution of the phase difference $\theta (t)$ (left) and
Josephson currents  (right) in the {\it ideal} BEC for the cases
a) and c) of Fig. 2. The exact $I(t)$ (solid bold curve ) and
approximate $\tilde{I}(t)$ (dotted curve) currents  are shown. In
b,d), the critical current $I_0$= 4.6 Hz is used to scale
$\tilde{I}(t)$. For $v^s_a$ (plot c), the chemical potential
difference is indicated. The barrier shift duration $T$=2 s is
marked by vertical dotted lines. For the reference, the zero line
is given in b,d). }
\end{figure}

\subsection{Analogy to Josephson effects}

Figure 6 shows evolution of the phase difference $\theta$ and
Josephson currents for the successful NLT of ideal BEC, presented
in Fig.2 a,c). Let's first consider the results for the soft
velocity profile $v_s(t)$ (Fig.6 c,d). They are less damaged by
dipole oscillations and so more convenient for the analysis. As
seen from (c), the phase difference $\theta$ starts from zero at
t=0, gets its maximum near the mid of the transfer (t=T/2=1000 ms)
and then decreases to the value $\theta_T \sim 0.027$. This
behavior roughly corresponds to the velocity profile, though the
final $\theta$ does not return to zero  but acquires a finite
value $\theta_T$. As shown below, the value of $\theta_T$ does not
depend on barrier velocity. So most probably this a geometric
phase accumulated during the NLT. For $t >T$, the  modest dipole
oscillations take place.

Since $\theta$ varies with time, we have here a phase-running
evolution, though with a small phase-locked ($\theta \approx
const$) region at $t\sim T/2$. In the first half of the evolution
($t<T/2$), the average chemical potential difference is $\Delta\mu
/h = \dot{\theta} \sim$ 0.043 Hz, i.e. is very small. The d.c.
assumes a constant phase difference $\theta$ and, therefore, zero
chemical potential difference $\Delta\mu$. The present process
demonstrates a small $\Delta\mu$ and so can be approximately
treated as a quasiadiabatic d.c.. The true d.c. takes place only
for shortly at the mid of the evolution ($t=T/2$).
\begin{figure*}
\begin{center}
\includegraphics[width=10cm]{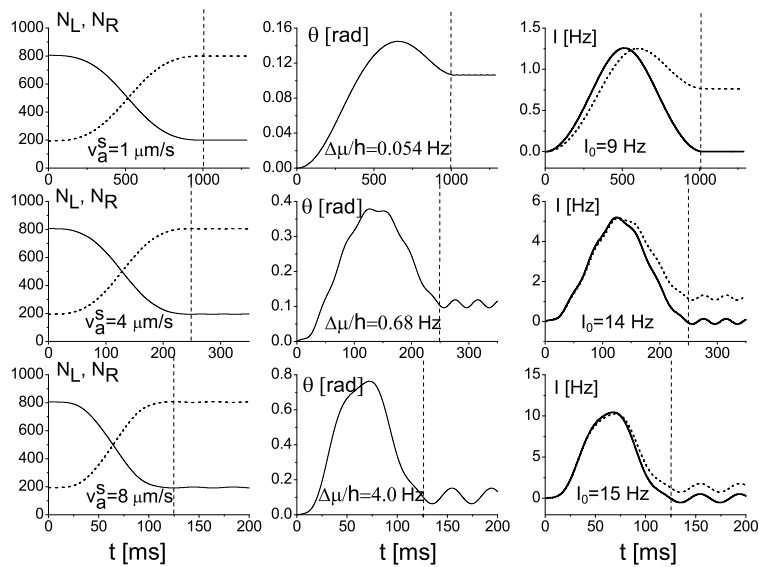}
\end{center}
\caption{Evolution of the {\it repulsive} BEC for the soft
velocity profile $v^s(t)$. Left plots: populations $N_L$ (solid
curve) and $N_R$ (dash curve). Middle plots: phase difference
$\theta$. Right plots: exact $I$ (bold solid line) and approximate
$\tilde{I}$ (dash line) Josephson currents. The slow ($v^s_a = 1
\; \mu$m/s, upper plots), middle ($v^s_a = 4 \; \mu$m/s, middle
plots) and fast ($v^s_a = 8 \; {\mu}m/s$, bottom plots) processes
are considered. For every case, the estimated chemical potential
difference $\Delta\mu /h$ and critical current $I_0$ are given.
The barrier shift durations are marked by vertical dash lines. }
\end{figure*}

Further insight to the process can be brought by a direct
inspection of Josephson currents. In Fig. 6d), the exact current
$I$  obtained through $\dot{z}$ and approximate current
$\tilde{I}$ determined through $\theta$ (see Eqs. (\ref{I}) and
(\ref{I_approx})), are depicted. For calculation of $\tilde{I}$,
the critical current $I_0$ = 4.6 Hz obtained from the condition
$I(t)=\tilde{I}(t)$ at $t=T/2$ is used (note that maximal $I <
I_0$). The plot d) shows that, for $t < T/2$, both $I$ and
$\tilde{I}$ are similar and closely follow the evolution of
$\theta$. Since $I(t) \propto \sin\theta$, we indeed have here a
Josephson-like phase-driven process.

For $t > T/2$, the behavior of $I$ and $\tilde{I}$ is different.
$I$  tends to zero (in accordance to Fig. 2c) while  the
approximate current $\tilde{I}$ approaches  a finite value (in
accordance to behavior of $\theta$ in Fig. 3c). The difference is
obviously caused by the final phase difference $\theta_T$.

In Fig. a,b), the same characteristics are presented  for the {\it
constant} velocity. Despite the average velocities of two profiles
are the same, $v^c_a=v^s_a=$6.4 nm s$^{-1}$, the evolution in
(a,b) is very polluted by dipole oscillations, which once more
shows the importance of using soft velocity profiles. In general,
up to the dipole oscillations, the behavior of $\theta$ and
currents in (a,b) is similar to those in (c,d). At the same time,
the plot b) provides an additional information: it shows that the
Josephson current $I$ is not constant even for the constant
velocity profile. So, in contrast to the statement \cite{Gi00},
the Josephson current is not necessarily proportional to the
barrier velocity .

In Figures 7 and 8, the evolution of the relevant characteristics
for the {\it repulsive} BEC is presented. Since the velocity
profile $v_c(t)$ leads to dipole oscillations which complicate the
analysis, we will further inspect only the soft profile $v_s(t)$.
In Fig. 7, the slow ($v^s_a = 1 \; \mu$m/s), middle ($v^s_a = 4 \;
\mu$m/s) and fast ($v^s_a = 8 \; {\mu}m/s < v_{\rm{crit}}$)
evolutions are considered. In all the cases, a successful NLT
takes place (left plots). The behavior of $\theta$ and currents in
the repulsive BEC is qualitatively similar to those  for the ideal
BEC. The main difference is in a significant enhancement of the
process rates. In particular, as mentioned above, the average
barrier velocities become 3 orders of magnitude bigger than for
the ideal BEC. The final phase difference $\theta_T$ remains
constant with increasing $v^s_a$. Its relative impact, being
decisive for a low velocity, becomes less important for large
velocities. It seems that just $\theta_T$ leads to some variance
of $I_0$. For a large $v^s_a = 8 {\mu}$m/s, we still have $\theta
< \pi/2$ and $I <I_0$. The chemical potential difference yet
remains modest, $\Delta\mu/h \sim $4.0 Hz, So, in general
agreement with the prediction \cite{Gi00}, this NLT can be
approximately treated as a quasiadiabatic phase-driven d.c.-like
process.

In Figure 8, the NLT near $v^{\rm{crit}}_a$ is considered (for the
soft velocity profile, this {\it average} critical velocity is
twice smaller than the {\it maximal} critical velocity in Figs.
4-5). It is seen that at the interval  $11 \; {\mu}m/s < v^s_a <
12 \; {\mu}m/s$ there is a pronounced transition to the a.c.-like
regime. For $v^s_a \ge 12 \; {\mu}m/s$, $\theta$ acquires a linear
time dependence while the current starts to oscillate with the
frequency $\omega \approx \Delta\mu/h$. The value of $\Delta\mu$
becomes much larger than for $v^s_a < v^{\rm{crit}}_a$. The
approximate current $\tilde{I}$ converges to the supercurrent $I$,
while the later approaches the critical current $I_0$. Altogether
all these factors unambiguously indicate the a.c. nature of the
final state. The high-frequency a.c. is modulated by low-frequency
dipole oscillations. The a.c. looks like MQST
\cite{Sme97,Rag99}(running phase, nonzero average population
imbalance $\langle z\rangle$) near the critical point ($v^s_a = 12
\; {\mu}m/s$) but deviates from MQST ($\langle z\rangle \to$ 0) at
higher velocities.
\begin{figure*}
\begin{center}
\includegraphics[width=10cm]{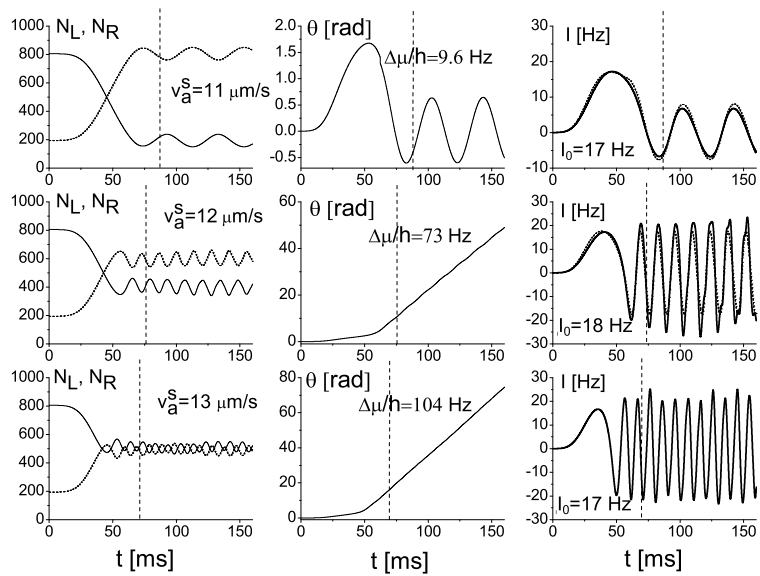}
\end{center}
\caption{Evolution near the critical velocity. The same as in Fig.
7 but for average velocities $v^s_a = 11 \; \mu$m/s, (upper
plots), $v^s_a = \; 12 \mu$m/s (middle plots), and fast $v^s_a =
\; 13 {\mu}m/s$ (bottom plots). }
\end{figure*}

Altogether, our analysis confirms that the NLT can be
approximately treated as a phase-driven, quasiadiabatic, d.c.-like
process occurring at $v < v^{\rm{crit}}$. For higher velocities $v
> v^{\rm{crit}}$, the NLT breaks down and transforms to a.c.
Note that the d.c. treatment of NLT  should be taken with a care.
Indeed, our calculations show that, for $v < v^{\rm{crit}}$, the
phase difference $\theta$ is not constant and the chemical
potential difference $\Delta\mu$ is not zero. Only smallness of
$\dot{\theta}$ and thus $\Delta\mu$ permits the d.c. treatment.

It should be emphasized that a cornerstone of d.c. in a weakly
coupled phase-driven system is an adiabatic following. Indeed, the
d.c. is adiabatic by definition (as a weak current yet unable to
produce quasiparticle excitations). Therefore, $v^{\rm{crit}}$ can
be treated  as a critical point for both d.c.$\to$ a.c.
\cite{Gi00} and  (quasi)adiabatic $\to$ nonadiabatic
\cite{Nest_JPB_09} transitions. Then, for example, the critical
velocity in quasiadiabatic Landau-Zener population transfer of the
repulsive BEC in a double-well trap \cite{Nest_JPB_09} can be
viewed both as a break of adiabatic following and as a d.c.$\to$
a.c. transition.

Finally note that, in the present study,  the trap is transformed
from the initial asymmetric form to the final opposite asymmetric
form, passing through the symmetric configuration at the mid of
the process (asym $\to$ sym $\to$ -asym transformation). Instead,
the previous theoretical \cite{Gi00} and experimental \cite{Le07}
studies used sym $\to$ asym and asym $\to$ sym transformations,
respectively. Despite these differences, the Josephson physics
behinds the evolutions is essentially the same. However, as
compared to \cite{Gi00,Le07}, our analysis is more complete in the
sense that i) the nonlinear impact is explored in detail and ii)
the crucial ingredient of the Josephson effects, the phase
difference, is numerically inspected.

\section{Summary}
\label{sec:summary}

The linear and nonlinear transport of BEC in a double-well trap
was investigated within the time-dependent three-dimensional
Gross-Pitaevskii equation, in close reference to parameters of
Heidelberg experiments \cite{Albiez_exp_PRL_05,Gati_APB_06}. The
calculations are performed for the total order parameter, thus
avoiding typical (two-mode, etc) approximations. The population
transfer is driven by a time-dependent barrier shift with a sharp
(rectangular) and soft ($\sim \cos^2(\omega t)$) velocity
profiles. It is shown that using the soft profile is crucial to
avoid strong dipole oscillations which significantly pollute the
transport and complicate its theoretical analysis and experimental
observation \cite{Le07}.

The calculations confirm our previous findings (obtained in the
simplified model \cite{Nest_JPB_09}) that repulsive interaction
between BEC atoms (and related nonlinearity of the problem)
significantly supports the NLT, making it possible in a wide
interval of barrier velocities. As compared to the ideal BEC, the
process can be three orders of magnitude faster. Besides, the
nonlinearity allows to produce the transport between stationary
states of essentially anisotropic trap. All these factors should
facilitate experimental investigation of NLT.

Note that the interaction effect is mainly caused by the rise of
the chemical potential. Hence the effect should depend on the
barrier form, being strong for smooth barriers whose penetrability
increases with the excitation energy and suppressed for sharp
barriers with a slight energy dependence of the penetrability.

Further, the relation of NLT and d.c. Josephson effect was
inspected in detail. As compared to previous studies
\cite{Za98,Rag99,Gi00,Meier01,Sak02}, the evolution of the phase
difference $\theta$ (a crucial ingredient of the Josephson effect)
was numerically explored. It was shown that, in accordance to
\cite{Gi00,Le07}, the NLT indeed can be approximately treated as
the d.c.. Above the critical  barrier velocity $v^{\rm{crit}}$,
the NLT decays into the a.c.. Note that the d.c. treatment of NLT
is actually an approximation because in NLT the phase difference
$\theta$ is not constant and the chemical potential difference
$\Delta\mu$ is not zero, which contradicts the d.c. definition.
However, because of the smallness of $\dot{\theta}$ and
$\Delta\mu$, the d.c. treatment is still reasonable.

The behavior of the transport near the critical velocity
$v^{\rm{crit}}$ was investigated in detail.  It is shown that
$v^{\rm{crit}}$ marks both d.c.$\to$ a.c.  \cite{Gi00} and
(quasi)adiabatic $\to$ nonadiabatic \cite{Nest_JPB_09}
transitions. These results emphasize an adiabatic nature of d.c.
in Bose-Josephson junctions (BJJ). Actually we deal here with a
general phase-driven adiabatic following of weakly-bound
two-component system. In this sense, a variety of (quasi)adiabatic
population transfer protocols (from familiar Landau-Zener
\cite{Lan,Zen} scheme and its generalizations \cite{Nest_JPB_09}
to modern adiabatic prescriptions \cite{Berry_09}) in internal and
external BJJ can be roughly considered as manifestations of the
d.c. Josephson effect.

\ack The work was partly supported by the RFBR grant 14-02-00723
and grants of University of Paul Sabatier (Toulouse, France. We
thank Prof. D. Gu´\'ery-Odelin for useful discussions.

\end{document}